\documentclass[twocolumn,aps,prl,showpacs,amsmath,amsfonts,amssymb,floatfix]{revtex4}

\usepackage{graphicx}
\usepackage{dcolumn}
\usepackage{bm}
\usepackage{epsfig}

\newcommand{\ZM}{\mathbb Z}

\newcommand{\ukr}{\hat U_{\text{\tiny{KR}}}}
\newcommand{\vkr}{\hat V_{\text{\tiny {KR}}}}
\newcommand{\uqkr}{\hat{\cal  U}_{\text{\tiny{KR}}}}
\newcommand{\voqkr}{{\cal  V}_{\text{\tiny{KR}}}(\theta,\varphi)}
\newcommand{\vqkr}{\hat{\cal V}_{\text{\tiny{KR}}}}
\newcommand{\vgtm}{\hat{\cal V}_{\text{\tiny{GTM}}}}

\newcommand{\ugtm}{\hat{\cal U}_{\text{\tiny{GTM}}}}

\usepackage{amsmath,amsfonts,amssymb,color,bbm}

\begin{document}
\title{Classical dynamical localization}

\author{Italo Guarneri$^{1,2}$, Giulio Casati$^{1,3,4}$, Volker Karle$^5$}
\affiliation{$^1$ Center for Nonlinear and Complex Systems,
Universit\`a degli Studi dell'Insubria, Via Valleggio 11, 22100 Como, Italy}
\affiliation{$^2$ Istituto Nazionale di Fisica Nucleare, Sezione di Pavia,
via Bassi 6, 27100 Pavia, Italy}
\affiliation{$^3$ International Institute of Physics, Federal University of Rio Grande do Norte, Natal, Brasil}
\affiliation{$^4$ CNISM and Istituto Nazionale di Fisica Nucleare, Sezione di Milano, via Celoria 16, 20133 Milano, Italy }

\affiliation{$^5$ Albert-Ludwigs Universit\"at Freiburg, D-79104 Freiburg im Breisgau, Germany}

\date{\today}
\begin{abstract}
We consider classical models of the kicked rotor type, with piecewise linear kicking potentials designed so that momentum changes only by multiples of a given constant. Their dynamics display quasi-localization of momentum, or quadratic growth of energy, depending on the arithmetic nature of the constant.  Such {\it purely classical}
features
mimic paradigmatic features of the {\it quantum} kicked rotor, notably dynamical localization  in momentum, or  quantum resonances. We present a heuristic explanation,  based on a classical phase space generalization of a well
known argument, that maps the quantum kicked rotor on a tight-binding model with disorder. Such results suggest  reconsideration of generally accepted views, that dynamical localization and quantum resonances are a pure result of quantum coherence.
\end{abstract}
\maketitle

Dynamical localization in the quantum kicked rotor (QKR) is a prototypical  example of how quantization can drastically modify  the qualitative features of classical chaotic motion\cite{loc0,Fish,Fx}. It is  assimilated to the Anderson localization in disordered solids\cite{loc1,Fish}, and, like the latter, it is considered to be an effect of quantum interference.  Dynamical localization is also expected  with kicked particles moving in a line. Experimental observations  on kicked cold atoms\cite{RZ} support such expectations, which are all the more natural, because  quantum kicked rotors  and quantum kicked particles are closely connected by Bloch theory \cite{FGR}: spatial periodicity enforces conservation of quasi-momentum, and the dynamics at fixed quasi momentum are those of a generalized QKR\cite{Da}.
Thus a crucial difference between the quantum and the classical dynamics of kicked particles is immediately apparent: notably, the former have a constant of the motion, while the latter have none.
In this paper we submit that exactly this difference plays a major role in the dynamical localization effect. We base on numerical results on a family of classical dynamical systems, which are subject to a purely classical conservation law for quasi-momentum.
These are models of the kicked rotor (KR) type, where the kicking potentials is a piecewise, continuous,  $2\pi$-periodic, function so that kicks can change  momentum only by integer multiples of a constant $\eta>0$.
They will be termed Generalized Triangle Maps (GTM) because they include as a particular example the Triangle map\cite{gc,gang}, that despite its formal simplicity still challenges exact analysis.

In this connection we'd like to recall a seminal paper\cite{Boris} in which the question was raised, whether the quantum inhibition of classical chaotic effects could be somehow explained by a  "discreteness of quantum phase space".  Following this idea, the classical kicked rotor was artificially discretized, and a limitation of the chaotic diffusion was observed.  Classical maps of the same kind  have later been derived in  a different way in ref.\cite{Berman}, and named 'classical models of quantum stochasticity'. These \emph{ad hoc} discretized models have been confirmed  to reproduce some quantum effects, like resonances, and limitations of chaotic diffusion.

Here we provide empirical and analytical evidence that purely classical models described by GTMs, indeed offer an intriguing imitation of the QKR. Our numerical results show that for strongly irrational $\eta/(2\pi)$, the KR diffusion is replaced by localization, or by ``quasi-localization", {\it i.e.} very slow   (power-logarithmic) transport in momentum space. If averaged over quasi-momenta, the quasi-localized momentum distributions display a clean exponential shape. For rational $\eta/(2\pi)$,
 and quasi-momentum commensurate to $2\pi$, quadratic growth of energy is observed, similar to  the QKR resonances\cite{Da}.
  \\``Quasi"-localization, instead of strict localization as in QKR, is likely to be due to absence of interference in GTM. Each kick changes the QKR state to a new state, where jumps by different multiples of $\hbar$ are coherently superposed and interfere in  ways that have no counterpart in GTM. Thus, in spite of general parallelism,  differences still  exist between the GTM  and the QKR dynamics, that reflect the fundamental difference of classical and quantum mechanics.  That  such differences leave room for classical lookalikes of dynamical localization and resonances  suggests re-assessment of commonplace views, that the latter effects  are pure manifestations  of quantum coherence.\\
Proper explanation of such numerical results demands  a purely classical analysis of the GTM. An exact argument to be reviewed later  shows that GTM resonances  have exactly the same origin as the QKR resonances, notably, conservation of quasi-momentum, together with  translation invariance (in momentum space). For quasi-localization, we instead present an heuristic argument that, somewhat paradoxically, is of a quantum origin. It is based on a construction due to Fishman, Grempel, and Prange (FGP)\cite{loc1,Fish, shep} that maps the QKR on a 1D tight-binding model with pseudo-random disorder. We first reformulate that  mapping in the Weyl phase space representation of quantum mechanics. The QKR is  thereby turned into a 2D tight-binding model with short-range hopping amplitudes and on-site potentials that are periodic at resonances and pseudorandom for sufficient incommensuration; in the latter case, exponential localization is inferred. The two directions in the lattice respectively correspond to momentum and to the harmonics of position. Next we note that, thanks to the special features of GTM,  the very same   construction  can be used to map the unitary Perron-Frobenius (P-F) operator of GTM  on a 2D lattice  model. Pseudorandomness is  the same as for the QKR and couplings are still short-range  in the direction of momentum; however, in the other direction they are now  long-range. The same line of reasoning  that so successfully works in the QKR case, now leads to predict localization in momentum, and de-localization in the harmonics of position.
   \begin{figure}

\epsfxsize7cm\epsfysize5cm\epsffile{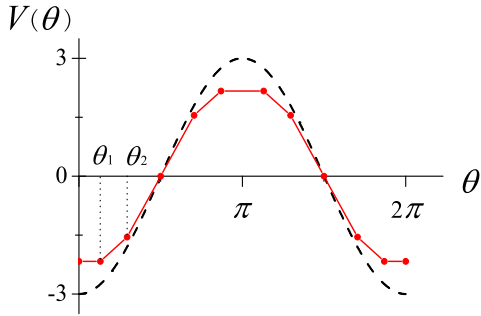}

\caption{The periodic potential $V(\theta)$ (red full line) is linear in between any two subsequent points $\theta_n$ where $\mu\sin(\theta_n)/\eta$ takes integer values (only those in $[0,\pi/2]$ are shown). Its piecewise constant slope is $V'(\theta)=j(\theta)\eta$ where  $j(\theta)$ is the integer that rounds $\mu\sin(\theta)/\eta$ toward $0$;  here $\mu=3,\eta=1.2$.
 In the limit of vanishing $\eta$, $V(\theta)$ converges to the standard kicked rotor potential  (dashed curve).
 }
 \label{dist4}
 \end{figure}
  \begin{figure}
 \begin{center}
\includegraphics[width=8cm]{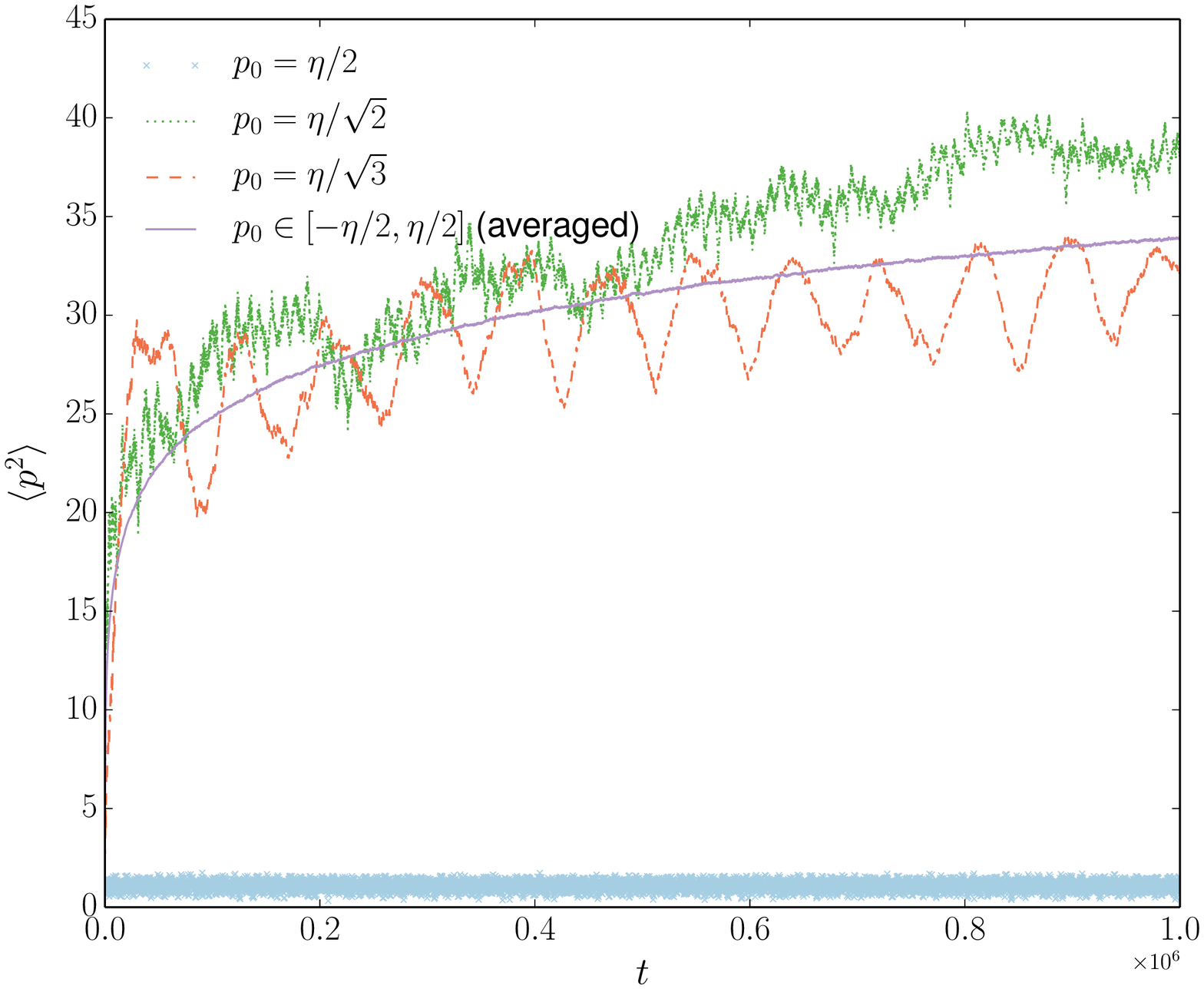}
\includegraphics[width=8cm]{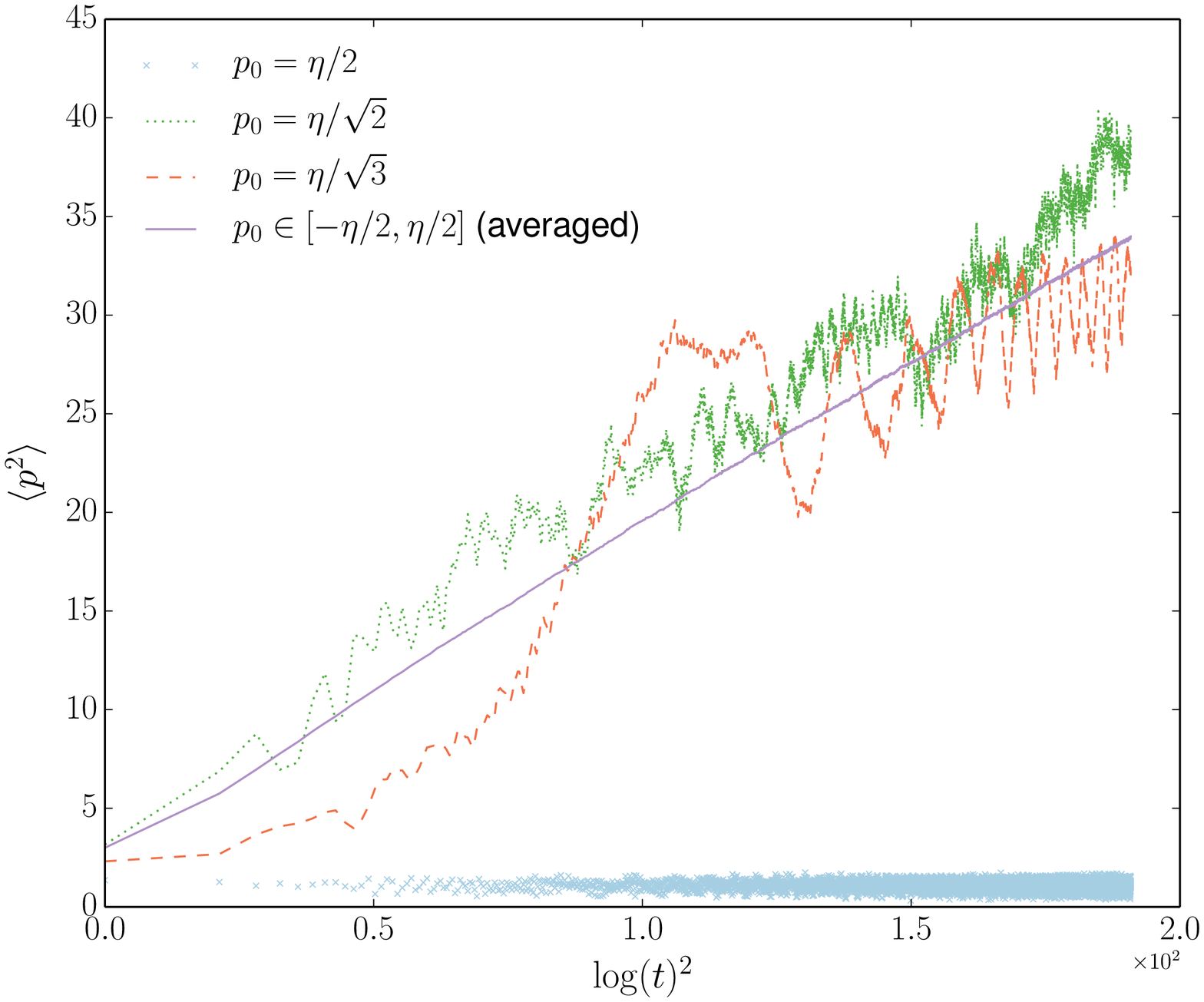}

\end{center}
\caption{$\langle p^2 \rangle$ of GTM  averaged over an ensemble of $10^6$  initial points with
$\theta_0$ randomly distributed in $(0,2\pi)$, versus time $t$ (upper plot) and versus $\log(t)^2$ (lower plot). Here  $\eta=\pi/gm$
($gm$ : the Golden Mean), $\mu=3$, and
$p_0=\eta/2, \eta/\sqrt{2}, \pi/\sqrt{3}$. Blue lines represent averages over  $5\times 10^6$ initial points $(\theta_0,p_0)$ randomly distributed in $(0,2\pi)\times(-\eta/2,\eta/2)$. All curves were averaged over $100$ iterations.}\label{spread1}
\end{figure}
  \begin{figure}
 \begin{center}
\includegraphics[width=8cm]{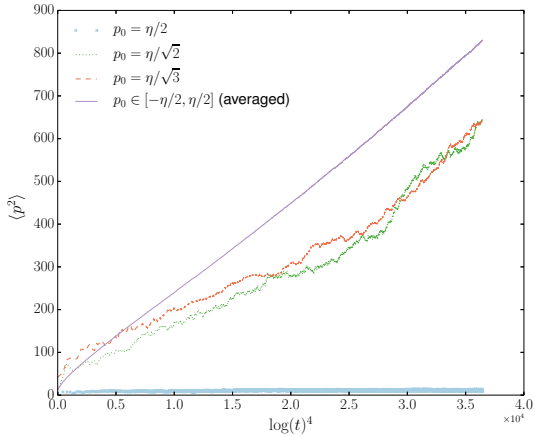}
\end{center}
\caption{Same as Fig.2 (lower)
but with $\mu=4$}
\label{spread1bis}
\end{figure}

\begin{figure}
 \begin{center}
\includegraphics[width=8cm]{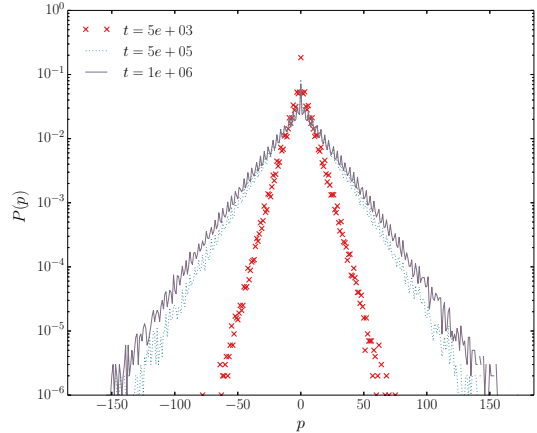}
\end{center}
\caption{Momentum distributions for ensembles of $5\times 10^6$ initial points $(\theta_0,p_0)$ randomly distributed in $(0,2\pi)\times(-\eta/2,\eta/2)$, $\mu=4$, and $\eta/\pi=gm$. The curves are averaged over the last $100$ kicks. }
\label{dist1}
\end{figure}

The GTMs we consider in this paper are  strictly classical  maps of the form
\begin{equation}
\label{map!}
p_{t+1}\;=\;p_t\;+\;V'(\theta_t)\;,\;\;\;\;\theta_{t+1}\;=\;\theta_t\;+p_{t+1}\;,
\end{equation}
where $V(\theta)$ is a  continuous, $2\pi$-periodic, piecewise linear potential, such that the possible values of $V'(\theta)$ ("channels") are a finite set of multiples of a constant $\eta>0$, and $V(\theta+\pi)=-V(\theta)$. In this paper we choose $V(\theta)$ as illustrated in fig (1). The map may also be read as a dynamical system in the 2-torus. Except for the fact that they are not chaotic, very little is known about such toral maps, even in the case when $V'(\theta)$ only takes two values\cite{kap}\cite{gang}. In that case,  the map is similar to a "triangle map" that describes the motion of a point mass in a right-triangular billiard.  Ergodicity of
 such billiards is a long standing issue, and in a recent paper\cite{Casati} contrary indications in that respect have been surmised, based on observation of a phenomenon there dubbed "exponential localization of invariant measures". It is easily proven that strict localization of  map (\ref{map!}) would forbid ergodicity of the corresponding toral map. A typical phase portrait of a toral map (\ref{map!}) is shown in the supplemental material. \\
 At all times $t$, $p_t=\beta+n_t\eta$, with $n_t$ integer and where the quantity $\beta\equiv$ mod$(p,\eta)$ is invariant under the map.
 The dynamics are thus described by the map
\begin{equation}
\label{map!!}
n_{t+1}\;=\;n_t\;+\;V'(\theta_t)/\eta\;,\;\;\;\;\theta_{t+1}\;=\;\theta_t\;+\;\beta\;+n_{t+1}\eta\;.
\end{equation}
 In our numerical investigation we have used GTMs with $3,5,7$ channels. Results are shown in Figs.\ref{spread1},\ref{spread1bis} and crucially depend on the arithmetics of the triple $\eta,\beta,\pi$.
With $\eta$ strongly incommensurate to $\pi$, and $\beta$ a low-order rational multiple of $\eta$,
 the average energy of ensembles with randomly generated $\theta_0$ undergoes strict localization (lowest curves). For $\beta$ incommensurate to $\eta$, we observe quasi-localization, {\it i.e.} slow, somewhat erratic  growth of energy  not faster than power-logarithmic, with an exponent that appears to depend on the number of channels. Additionally averaging over quasi-momentum yields  clean indications in this sense (full lines).
 The $\beta$-averaged, quasi-localized momentum distributions show a remarkably clean exponential decay away from a central tiny peak (Fig.\ref{dist1}).
  The time scale for the onset of such exponential distributions rapidly increases on decreasing $\eta$, as the initial spreading in momentum approaches the classical diffusion of the kicked rotor.  At small $\eta$  numerical analysis of the quasi-localized regime becomes a prohibitive computational task.
When $\beta$, $\eta$ and $\pi$ are mutually commensurate, ballistic momentum growth is observed, at least for sufficiently large values of $\mu/\eta$. If $\eta$ is identified with an effective Planck constant, this behavior reproduces the quantum resonances  of the generalized QKR\cite{Da} (except for the fact that the latter do not depend on the kicking strength).  This is explained as follows\cite{Berman}. Let $\beta$ and $\eta$ be commensurate: $\beta s=\eta r\equiv rs\lambda$, with $r,s$ coprime integers. Then, at all times $t$, $p_t=N_t\lambda$ and $\theta_t=\theta_0+M_t\lambda$ with $N_t,M_t$ integers. Replacing in (\ref{map!!}) a map is obtained for the integers $N_t$ and $M_t$:
\begin{equation}
\label{mapint}
N_{t+1}\;=\;N_t\;+\;\Phi(M_t)\;,\;\;\;M_{t+1}\;=\;M_t\;+\;N_{t+1}\;,
\end{equation}
where, for fixed $\theta_0$, $\Phi(M_t)\equiv\lambda^{-1}V'(\theta_0+M_t\lambda)$ is an integer valued quasi-periodic function, with  a finite number of values. If in addition $\eta$ (and hence $\lambda$) is commensurate to $2\pi$, $\lambda
=2\pi p/q$ with $p,q$ coprime integers, then $\Phi$ is periodic, and  map (\ref{mapint}) commutes with translations of both variables by multiples of $q$. Therefore it defines a map $\cal M$ of  the discrete $2$-torus ${\mathbb T}_q\times{\mathbb T}_q$ in itself, where ${\mathbb T}_q$ is the set of congruence classes mod$(q)$.  As $\cal M$  is bijective in a finite set, all its trajectories are periodic; so,  for all choices of $\theta_0$ and $p_0$ which are consistent with the given $\beta$, there are
a period $T$ and integers $K,L$ so that $\theta_{mT}=\theta_0+Kmq\lambda$ mod$(2\pi)$ and $p_{mT}=p_0+Lmq\lambda$ for all integer $m$. If the number of channels is sufficiently large,
some orbits have  $L>0$; along such orbits the momentum $p_t$ increases, in the average, proportional to $Lq\lambda t/T$. Quadratic  growth of the mean energy follows.
\\Next we show that a theoretical understanding about the purely classical GTM  can  be obtained from quantum localization theory.
To this end we consider the P-F operator $\ugtm$ for the GTM dynamics. For given $\beta$, the phase space of map (\ref{map!!}) is $\Omega={\mathbb T}\times{\mathbb Z}$ and $\ugtm$ unitarily acts on square-summable functions $\Psi\in L^2(\Omega)$ so that  $\ugtm\Psi(\theta_t,n_t)=\Psi(\theta_{t-1},n_{t-1})$. To the unitary operator $\ugtm$ we will associate a $2D$ lattice problem, implementing the FGP construction that was used\cite{loc1} to map the quantum kicked rotor on a $1D$ lattice problem. Here we outline this calculation, leaving details for the supplemental material.  We first derive a phase-space version of the FGP  construction and to this end we  exploit the Weyl correspondence $\mathfrak W$, that maps Hilbert-Schmidt operators $\hat A$ (e.g., states) in the Hilbert space of the QKR to square-summable function $\Psi={\mathfrak W}(\hat A)$ on $\Omega$, according to:
\begin{gather}
\label{weyl1}
\Psi(\theta,n)\;=\;\tfrac1{\sqrt{2\pi}}\sum\limits_{l\in\ZM}\langle e_l|{\hat A}|e_{n-l}\rangle\;e_{2l-n}(\theta)\\
=\;\int_0^{2\pi}d\theta'\langle\theta'|{\hat A}|2\theta-\theta'\rangle\;e_n(\theta-\theta')\;,
\label{weyl2}
\end{gather}
where $e_l(\theta)=(2\pi)^{-1/2}\exp(il\theta)$. As ${\hat A}$ evolves into ${\hat U}{\hat A}{\hat U}^{\dagger}$ where $\hat U$ is some unitary evolution operator, its Weyl representative $\Psi$ evolves into $
{\hat{\cal U}}\Psi\;=\;{\mathfrak W}\bigl(\hat U{\mathfrak W}^{-1}(\Psi){\hat U}^{\dagger}\bigr)\;.
$. This defines the unitary propagator  $\hat{\cal U}$ in $L^2(\Omega)$ that describes  evolution in the Weyl representation. Let in particular $\hat U$ be the Floquet operator $\ukr$  of  the QKR with quasi-momentum $\beta$: $\ukr=\exp(-i\vkr)\exp(-i\hat T)$, where:
\begin{equation}
\label{genrot}
V_{\text{\tiny KR}}=\hbar^{-1}\mu\cos(\theta)\;,\;\;\;\hat T=\hbar^{-1}(-i\hbar\tfrac d{d\theta}+\beta)^2/2\;.
\end{equation}
A calculation based on (\ref{weyl1}) and (\ref{weyl2}),  we find that:
\begin{equation}
\label{prodkr}
\uqkr=\exp (-i\vqkr)\exp(-i\hat {\cal T})\;,
\end{equation}
where:
\begin{gather}
\label{rrot}
\hat{\cal T}\;=\;-i\bigl(\tfrac12\hbar n\;+\;\beta\bigr)\frac{d}{d\theta}\;,\\
\mathfrak F\vqkr\Psi(\theta,\varphi)\;=\;
-2\mu\hbar^{-1}\sin(\theta)\sin(\varphi)\;{\mathfrak F}\Psi(\theta,\varphi)\;.
\label{kck}
\end{gather}
where $\mathfrak F$ denotes the Fourier transform :
${\mathfrak F}\Psi(\theta,\varphi)\equiv\sum_{n}\Psi(\theta,n) e_n(\varphi)$. At this point we come  to the FGP construction . In its original version\cite{Fish, loc1}, it maps the
QKR eigenvalue equation $\ukr\psi=e^{i\omega}\psi$ on a 1D tight-binding problem. The equation is indeed equivalent to $({\hat W} +{\hat Z})\phi=0$, where $\hat W$ and $\hat Z$ are inverse Cayley transforms  of $\exp(-i\vkr)$ and $\exp[-i({\hat T}+\omega)]$ respectively, and $\phi=(i+{\hat Z})^{-1}\psi$. Now
$\exp(-i({\hat T}+\omega))$ (hence $\hat Z$) has a pure point spectrum,  $\exp(-i\vkr)$ (hence $\hat W$)  is invariant under translations over the eigenbasis of $\hat T$, and if $\mu<\pi$ then $W$ is bounded so, written in that basis,
$({\hat W} +{\hat Z})$  looks like a lattice Hamiltonian, where
 the eigenvalues of $\hat Z$ play the role of on-site potentials, and $\hat W$ describes hopping between sites. This construction  works unaltered if $\ukr$ is replaced by any  operator (in an arbitrary Hilbert space), that comes in the form of a product of two unitary operators with the above properties. For $\mu<\pi/2$ this is the case with $\uqkr$ thanks to (\ref{prodkr}),(\ref{rrot}),(\ref{kck}), and FGP immediately yields the following  2D lattice equation:
\begin{equation}
\label{shm}
\sum\limits_{n',k'\in\ZM}W_{n-n',k-k'}{\Phi}_{ n'k'}\;+\;Z_{nk}(\omega){\Phi}_{nk}\;=\;0\;,
\end{equation}
where $Z_{nk}(\omega)\equiv \tan(\chi_{nk}(\omega))$, $\chi_{nk}(\omega)=
[\omega-(n\hbar/2+\beta)k]/2$.  For $\mu<\pi\hbar/2$ the
couplings $W_{n,k}$ are the Fourier coefficients of the analytic function
$\tan(\mu\hbar^{-1}\sin(\theta)\sin(\varphi))$ so they decay exponentially fast, and
(
\ref{shm}) is formally similar to an eigenvalue equation for a 2D tight-binding model with short range hopping.  When $\hbar$ is strongly incommensurate to $2\pi$, the potential is pseudorandom, and exponential localization follows. At resonances $\hbar$ and $\beta$ are commensurate to $2\pi$, so the potential is periodic, enforcing  extended eigenfunctions, and ballistic propagation . \\
This construction  is not applicable as it is  when $\mu>\pi\hbar/2$, because then $\tan(\mu\hbar^{-1}\sin(\theta)\sin(\varphi))$ has non-integrable singularities. This difficulty  is circumvented by an improved method\cite{shep}. For the Weyl representation of QKR, this method  replaces (\ref{shm}) by:
\begin{equation}
\label{dima}
\sum\limits_{n',k'}|{\tilde W}_{n-n',k-k'}|\sin(\chi_{n'k'}(\omega)+\phi_{n-n',k-k'}){\tilde\Phi}_{n'k'}\;=\;0\;,
\end{equation}
where ${\tilde W}_{n,k}$ are the Fourier coefficients of $e^{-i\voqkr/2}$, and $\phi_{n,k}$  are their phases \footnote{The selfadjoint Hamiltonian underlying (\ref{dima}) is (in operator form)\cite{shep}:
${\hat H}=\cos(\tfrac12{\vqkr})\tan(\tfrac12\omega-\tfrac12\hat{\cal T})\cos(\tfrac12\vqkr)-\tfrac12\sin(\vqkr).
$}
 In this formulation, disorder also appears in couplings, which still decay exponentially fast.\\
 We've thus rephrased the FGP construction for the QKR in the phase-space representation. This was possible, thanks    to a special structure of the Weyl propagator of the QKR, as a unitary  operator in $L^2(\Omega)$.
 Now we'll show that the same is true  with the completely classical P-F operator of GTMs. To see this, just replace $\hbar$ by $\eta$ throughout, and let the prefactor of ${\mathfrak F}\Psi$ on the rhs of (\ref{kck}) be replaced by $2\varphi V'(\theta)/\eta$. Then, instead of $\vqkr$, (\ref{kck}) defines a new operator $\vgtm$, and it is easily seen that:
\begin{gather}
\label{repla}
e^{-i\vgtm}\Psi(\theta,n)\;
=\; \Psi(\;\theta, n-2V'(\theta)/\eta\;)\;.
\end{gather}
Using (\ref{prodkr}) and (\ref{rrot}), the full propagator  $\uqkr$ is replaced by:
\begin{gather}
\ugtm\Psi(\theta,n)\;=\;\Psi(\theta',n')\;,\nonumber\\
\theta'=\theta-\eta n/2-\beta,\;\;\;n'=n-2V'(\theta')/\eta\;,
\label{pf}
\end{gather}
Restricting  to even values $n$, and rescaling  $n$ by $1/2$, the map in (\ref{pf}) is  the inverse of the reduced GTM map (\ref{map!!}),
so $\ugtm$ is the P-F operator for the dynamical system (\ref{map!!}). This  opens the way to mapping on a 2D lattice model. $\vqkr$ has to be replaced by
$\vgtm$, so the improved formulation (\ref{dima}) is necessary,  because $\tan(\varphi V'(\theta)/\eta)$ has non-integrable singularities.``Disorder" is  the same, but couplings are different:
$$
{\tilde W}_{n-n',k-k'}\;=\;\tfrac1{2\pi}\int_{I_{n-n'}}d\theta\;e^{-i(k-k')
\theta}\;,
$$
where $I_n$ is the interval wherein $2V'(\theta)=n\eta$. In the $n$-direction (momentum) such couplings vanish whenever $|n-n'|\eta$ is larger than the maximum of $|V'(\theta)|$.  In the $k$-direction (harmonics of position) they slowly decay proportional to $|k-k'|^{-1}$ due to discontinuities  of $V'(\theta)$. On such grounds, whenever $\eta$ is incommensurate to $\pi$  we are led to expect (i) localization
in momentum, and (ii) de-localization over the harmonics of position. We consider (i) to be consistent with numerical results, because the argument is too crude to discriminate quasi-localization from strict localization; inferring  the power-logarithmic spreading from the  2D lattice dynamics is a  nontrivial interesting problem. At variance with QKR, (ii) implies a continuous GTM spectrum in all cases. In order to check (ii),  we have numerically computed the P-F evolution of a given function of $n$ and $\theta$. Fourier expansion at each time $t$ yields amplitudes $f_{nk}(t)$ at the sites in the 2D lattice. Our numerical results show that the distribution $P(k,t)\equiv\sum_n|f_{nk}(t)|^2$  rapidly spreads over the whole available Fourier basis.\\
Our present evidence of GTM quasi-localization and resonances is for cases when $\eta/\pi$ is either strongly irrational (in fact, equivalent to the Golden Mean), or rational. A better understanding of this intriguing dynamical behavior will require analysis of how it depends on the degree of irrationality  of $\eta/\pi$ .

\end{document}